\documentclass[a4paper]{article}

\usepackage[english]{babel}
\usepackage[utf8]{inputenc}
\usepackage[T1]{fontenc}
\usepackage[top=3.2cm, bottom=3.2cm, left=2.5cm, right=2.5cm]{geometry}

\usepackage[round]{natbib}
\usepackage{amsmath}
\usepackage{bm}
\usepackage{fix-cm}

    \newcommand{\vect}[1]{\bm{#1}}
    
    \newcommand{\partiald}[2]{\frac{\partial #1}{\partial #2}}
    \newcommand{\n}{\boldsymbol{\nabla}}

    \DeclareMathSymbol{d}{\mathalpha}{operators}{`d}

    \DeclareMathAlphabet{\mathsfit}{T1}{\sfdefault}{\mddefault}{\sldefault}
    \SetMathAlphabet{\mathsfit}{bold}{T1}{\sfdefault}{\bfdefault}{\sldefault}
    \newcommand{\mathsfbi}[1]{\bm{\mathsfit{#1}}}

    \numberwithin{equation}{section}

\usepackage{titling}
\pretitle{
  \begin{center}
    \Large\bfseries\scshape
}
\posttitle{%
  \end{center}%
}
\usepackage{authblk}

\usepackage{graphicx}
\usepackage[margin=0pt,
            font=small,
            labelfont=bf,
            labelsep=endash]{caption}
\usepackage{wrapfig}
\usepackage{subcaption}

\usepackage[calcwidth]{titlesec} 
\titleformat{\section}[block]{\large\centering\scshape\thesection. }{}{0em}{}[]
\titleformat{name=\section,numberless}[block]{\large\centering\scshape}{}{0em}{}[]
\usepackage{xcolor} 
\usepackage{hyperref} 
\hypersetup{
	colorlinks,
	citecolor=blue,
	filecolor=black,
	linkcolor=blue,
	urlcolor=black
}

\usepackage{tikz}
\usetikzlibrary{arrows.meta}
\usetikzlibrary{decorations.pathreplacing,decorations.markings}

\title{Numerical study of a viscous breaking water wave and the limit of vanishing viscosity}

\date{}

\author[,1]{Alan Riquier\thanks{Alan.Riquier@ens.fr}}
\author[,1]{Emmanuel Dormy\thanks{Emmanuel.Dormy@ens.fr}}

\affil[1]{Département de Mathématiques et Applications, CNRS UMR-8553, École Normale Supérieure, PSL University, 75005 Paris, France.}

\begin{document}

\maketitle

\begin{abstract}
We introduce a numerical strategy to study the evolution of {2D} water waves in the presence of a plunging jet. The free-surface Navier-Stokes solution is obtained with a finite but small viscosity. We observe the formation of a surface boundary layer where the vorticity is localised. We highlight convergence to the inviscid solution. The effects of dissipation on the development of a singularity at the tip of the wave is also investigated by characterising the vorticity boundary layer appearing near the interface.
\end{abstract}

\section{Introduction}
\label{sec:intro}

Wave breaking is among the most common and probably most beautiful fluid flow occurring in nature. Yet, it remains extremely challenging to study from a modeling point of view. Being a strongly non-linear  phenomenon, usual analytical methods fail to capture the full mechanisms. Also, the different scales involved, as well as the free surface flow, significantly complicate any numerical approach. Our purpose is to provide a new framework to compute numerical viscous water waves, allowing the free-surface to overturn until the point at which the interface intersects itself.

Remarkably, the inviscid water wave problem can be reformulated using quantities defined on the water-air interface only \citep{Zakharov1968}. Several numerical developments are based on such approaches \citep{LonguetHigginsCokelet1976,NewPeregrine1985,VinjeBrevig1981,Baker82,baker_xie_2011}.
Even though a complete mathematical description of breaking waves seems unlikely, thorough partial analytical studies highlighted the possibility of self-similar solutions of a hyperbolic crest leading to a singularity \citep{LonguetHiggins1982,New1983}.

An alternative route to study the wave breaking phenomenon, consists in considering the free-surface Navier-Stokes equations with a small, but finite, viscosity \citep{Chen1999,Raval2009,Mostert2022,DiGiorgio2022}. Various instabilities, including the formation of aerated vortex filaments after the breaking stage could thus be described \citep{lubin_glockner_2015,Lubin2019}. 
The difficulty then relies in approaching the relevant limit of vanishing viscosity and surface tension (when it is included). Surprisingly, none of the recent studies encompassing the viscosity has compared their results with the inviscid case.

Here we introduce a finite-element formulation for the free-surface Navier-Stokes flow and investigate a plunging breaker case. {We investigate the formation of a sharp tip on the plunging breaker.} We achieve convergence to the Euler solution \citep{DormyLacave2023} and study the role of the viscous boundary layer in regularising the interface.

\section{Mathematical formulation}
\label{sec:MathAndNumerics}

\begin{figure}
    \centering
    \begin{tikzpicture}[scale=2.1]
        \fill[color=cyan!20]
                (0,0) -- (0,1.5) 
                -- plot [domain=0:2*pi,samples=200] (\x,{1+0.5*cos(\x r)}) 
                -- (2*pi,0) -- cycle;
        \draw[->] (0,0) -- (2*pi+0.25,0) node[below right]{$x$} ;
        \draw[->] (0,0) -- (0,1.75) node[above left]{$y$} ;
        \draw (2*pi,0) -- (2*pi,1.5) ;
        \draw [domain=0:2*pi,samples=200] plot (\x,{1+0.5*cos(\x r)});
        \draw[dashed] (0,1) -- (2*pi,1);
        \draw[|-|] (-0.1,0) -- (-0.1,1) node[midway,left]{$h_0$};
        \draw[->] (pi,0) -- (pi,0.475) node[midway,right]{$h_0 + a\cos(kx)$};
        \draw (1,0.5) node{$\Omega_{t=0}$};
        \draw (6,0.1) node{$\Gamma_b$};
        \draw (6,1.6) node{$\Gamma_{s,t=0}$};
        \draw[|-|] (0,-0.1) -- (2*pi,-0.1) node[midway,below]{$L$}; 
    \end{tikzpicture}
    \caption{Geometry of the initial ($t=0$) domain for the viscous water waves problem.}
    \label{fig:def}
\end{figure}
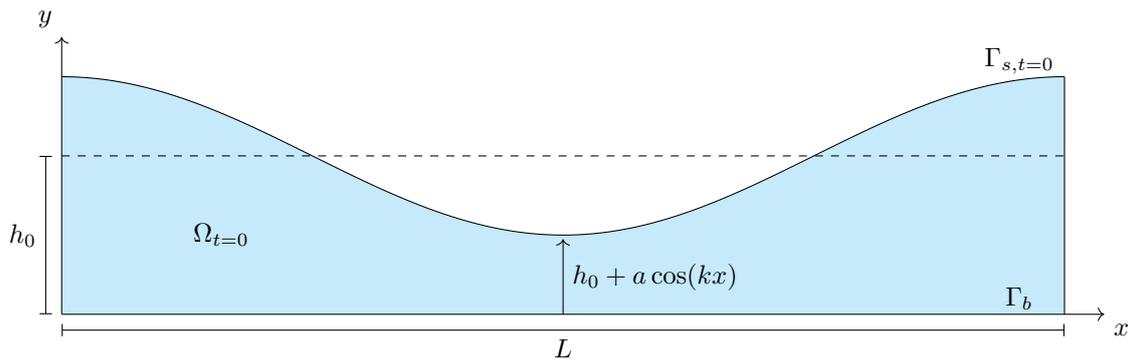

We consider the two-dimensional water domain $\Omega_t$ depicted in figure \ref{fig:def}, which we assume  $L$-periodic in $x$. Along the $y$ direction, $\Omega_t$ is encapsulated between a rigid bottom $\Gamma_b$ of the fluid domain and the water-air interface $\Gamma_{s,t}$, {represented by a time-dependent parametrised curve $\vect{\gamma}_t$ }. The former will remain unchanged
throughout the study whereas the latter is evolving with time. 
Nondimensional quantities are  defined using the height of fluid at rest $h_0$ as unit of length, $\sqrt{gh_0}$ as unit of velocity ($g$ being the gravitational acceleration), and $\rho g h_0$ as unit of pressure ($\rho$ being the fluid density, assumed to be homogeneous).
The Reynolds number is then defined as 
$\mathrm{Re} = {\rho h_0 \sqrt{gh_0}}/{\mu},$
where $\mu$ is the fluid dynamic viscosity, also assumed to be homogeneous. {Some earlier studies, \textit{e.g.} \citet{Chen1999, IAFRATI_2009, Deike2015, DiGiorgio2022, Mostert2022}, used the deep-water scaling to define their Reynolds number, $\mathrm{Re}_{\mathrm{dw}} = \rho \sqrt{g\lambda^3}/\mu$, where $\lambda$ is the wavelength of the initial wave, here set to $L\, ,$ thus $\mathrm{Re}_{\mathrm{dw}}=(L/h_0)^{3/2}\,\mathrm{Re}$.}

The incompressible Navier-Stokes equations in the domain $\Omega_t$ then take the form
\begin{equation}\label{eq:NS_normal}
    \left\{ \begin{array}{rcll}
        \cfrac{\partial\vect{u}}{\partial t} + \big(\vect{u}\cdot\n\big)\vect{u} +\n p - {\mathrm{Re}}^{-1}\n^2\vect{u} &=& -\hat{\vect{y}} & \mathrm{in}\ (0,T)\times\Omega_t \, ,\\
        \n\cdot\vect{u} &=& 0 & \mathrm{in}\ (0,T)\times\Omega_t \, ,\\
        \vect{u}\cdot\hat{\vect{n}} &=& 0 & \mathrm{on}\ (0,T)\times\Gamma_b\, ,\\
        \hat{\vect{t}}\cdot\mathsfbi{S}\cdot\hat{\vect{n}} &=& 0 & \mathrm{on}\ (0,T)\times\Gamma_b \, ,\\
        -p\,\hat{\vect{n}} + {2}{\mathrm{Re}}^{-1}\,\mathsfbi{S}\cdot\hat{\vect{n}} &=& {\kappa\, {\text{Bo}^{-1}} \,\hat{\vect{n}}} & \mathrm{on}\ (0,T)\times\Gamma_{s,t}\, ,\\
        \vect{u}(0,\cdot) &=& \vect{u}_0 & \mathrm{in}\ \Omega_0 \, .
    \end{array} \right. 
\end{equation}
\textit{Stress-free} boundary conditions are imposed at the water-air interface $\Gamma_{s,t}$, where $\kappa$ is the surface curvature and $\text{Bo}^{-1} = \sigma/ \rho g h_0^2 $ the Bond number (with $\sigma$ the surface tension). At the bottom $\Gamma_{b}\, ,$ we use stress-free in the tangential ($\hat{\vect{t}}$) direction and enforce no penetration in the normal direction.
The above formulation is sometimes referred to as `single fluid' as the air density has been dropped. $\sigma$ denotes the surface tension at the free-surface.
Finally, $\mathsfbi{S}$ denotes the \textit{stress tensor} defined as $\mathsfbi{S}(\vect{u}) = \frac{1}{2}\big( \n \vect{u} + ( \n \vect{u} )^t\big) \, .$

The interface $\Gamma_{s,t}$ being a material surface, a two-dimensional advection problem needs to be solved to follow the evolution of $\vect{\gamma}_t$ with time,
\begin{equation}\label{eq:advect_inter}
    \frac{\partial\vect{\gamma}_t}{\partial t} = \vect{u}\big(t,\vect{\gamma}_t\big).
\end{equation}
With this approach the shape of the interface does not need to be a graph. This is a key property to describe breaking waves and causes difficulties with several formulations of the water-wave problem \citep[see][chapter 1]{lannes2013}.

We consider as initial condition a simple wave (solution of the linearised equations) of finite amplitude $a$ so that the initial interface can be represented as
\begin{equation}
    \Gamma_{s,t=0} = \big\{ \big(x,h_0 + a\cos(kx)\big), \text{ with } x\in [0,L] \big\},
\end{equation}
where $k = 2\pi/L$ denotes the wave number.
The initial velocity $\vect{u}_0$ on the interface $\Gamma_{s,t=0}$ is given by a finite amplitude extension of the first-order two-dimensional solution of the inviscid water waves problem \citep[e.g.][chapter 2]{johnson},
\begin{equation}
    \vect{u}_0(x)= a\sqrt{gk\tanh(kh_0)}\cdot\begin{bmatrix}
		\big( \tanh kh_0 \big)^{-1}\cos kx\\\sin kx
	\end{bmatrix}.
\end{equation}
The initial velocity $\vect{u}(0,\cdot)$ in the full fluid domain could be approximated from the series expansion in $ka$. However, since we consider a large amplitude wave, we rather compute it numerically. This is easily achieved by solving the Laplace equation for the initial velocity potential $\phi_0$ in $\Omega_{t=0}$ (hence assuming a vanishing initial vorticity)
\begin{equation}
    \Delta\phi_0 = 0\ \mathrm{in}\ \Omega_{0} \qquad\text{so that}\qquad\vect{u}(t=0,\vect{x}) = \n\phi_0(\vect{x}),
\end{equation}
with boundary conditions
\begin{equation}
	\cfrac{\partial\phi_0}{\partial\hat{\vect{n}}} = 0\text{ on }\Gamma_b\qquad\text{and}\qquad
	\cfrac{\partial\phi_0}{\partial\hat{\vect{n}}} = \vect{u}_{0}\cdot\hat{\vect{n}}\text{ on }\Gamma_{s,t=0},
\end{equation}
corresponding to $\vect{u}(0,\cdot)\cdot\hat{\vect{n}} = 0$ on $\Gamma_b$ and $\vect{u}(0,\cdot)\cdot\hat{\vect{n}} = \vect{u}_{0}\cdot \hat{\vect{n}}$ on $\Gamma_{s,t=0}$.

In order to achieve a finite-elements formulation, we need to rewrite problem \eqref{eq:NS_normal} in a weak form. This is achieved by introducing the function space
\begin{equation}
    \mathbf{H}^1_{\Gamma_b}(\Omega_t) = \big\{ \vect{v}\in \big( H^1(\Omega_t) \big)^2\text{ such that } \vect{v}\cdot\hat{\vect{n}} = 0 \text{ on } \Gamma_b \big\},
\end{equation}
where $H^1(\Omega_t)$ stands for the usual Sobolev space. We then multiply the velocity equation by an arbitrary test function $\vect{v}\in\mathbf{H}^1_{\Gamma_b}(\Omega_t) $ and the incompressibility condition by another test function $q\in L^2(\Omega_t)$. Integrating over the whole domain $\Omega_t$ and making use of our boundary conditions \citep[e.g.][]{guermond2013} leads to the following iational formulation: find $\vect{u}\in \mathcal{C}^1\big([0,T);\mathbf{H}^1_{\Gamma_b}(\Omega_t)\big)$ and $p\in L^{\infty}\big([0,T);L^2(\Omega_t)\big)$ such that
\begin{equation}
    \begin{aligned}
    \label{eq:NS_weak}
    \int_{\Omega_t} \left[\vect{v}\cdot\frac{\partial\vect{u}}{\partial t} + \vect{v}\cdot\big(\vect{u}\cdot\n\big)\vect{u} \right.&+\left. {2}{\mathrm{Re}}^{-1}\, \mathsfbi{S}(\vect{v}) : \mathsfbi{S}(\vect{u}) - p\n\cdot\vect{v} - q\n\cdot\vect{u} - \vect{v}\cdot\vect{g}\right]\,d^2\vect{x} \\
    &= \int_{\Gamma_{s,t}} {\kappa}{\text{Bo}^{-1}} \, \vect{v}\cdot\hat{\vect{n}} \,dS
\end{aligned}
\end{equation}
and $\vect{u}(0,\cdot) = \vect{u}_0$ for all test functions $\vect{v}\in\mathbf{H}^1_{\Gamma_b}(\Omega_t)$ and $q\in L^2(\Omega_t)\, .$

\section{Numerical discretisation}
\label{sec:numerics}

We numerically approximate solutions of system \eqref{eq:NS_weak} using the FreeFEM language \citep{FreeFem}.

Once equation \eqref{eq:NS_weak} has been numerically integrated, the interface must be advected following \eqref{eq:advect_inter}. This is achieved using an Arbitrary Lagrangian Eulerian (ALE) method \citep{ALE_HIRT1974}. At each iteration, we numerically construct a mesh velocity solving 
\begin{equation}
    \label{eq:LaplaceMeshVeloc}
    \left\{\begin{array}{rcll}
        \Delta \vect{w} &=& 0 & \text{in\quad } \mathcal{T}_h^n \, ,\\
        \vect{w} &=& 0 & \text{on\quad} \Gamma_{b,h} \, ,\\
        \vect{w} &=& \vect{u}(t,\cdot) & \text{on\quad} \Gamma_{s,t_n,h} \, .\\
    \end{array}\right. 
\end{equation}
where the $h$ subscript denotes the discrete numerical boundary.
We then advect the mesh vertices using the mesh velocity $\vect{w}$. It is this necessary, in this approach, to recompute FE matrices at each time step since the mesh is advected (the Jacobian of the transformation between the reference element and the global cell is thus not constant). The $\vect{w}$ field is subtracted to
the advection term of the fluid equation.

\begin{figure}
    \centering
    \includegraphics[trim=1.7cm 9cm 0cm 9cm, clip, width=0.9\textwidth]{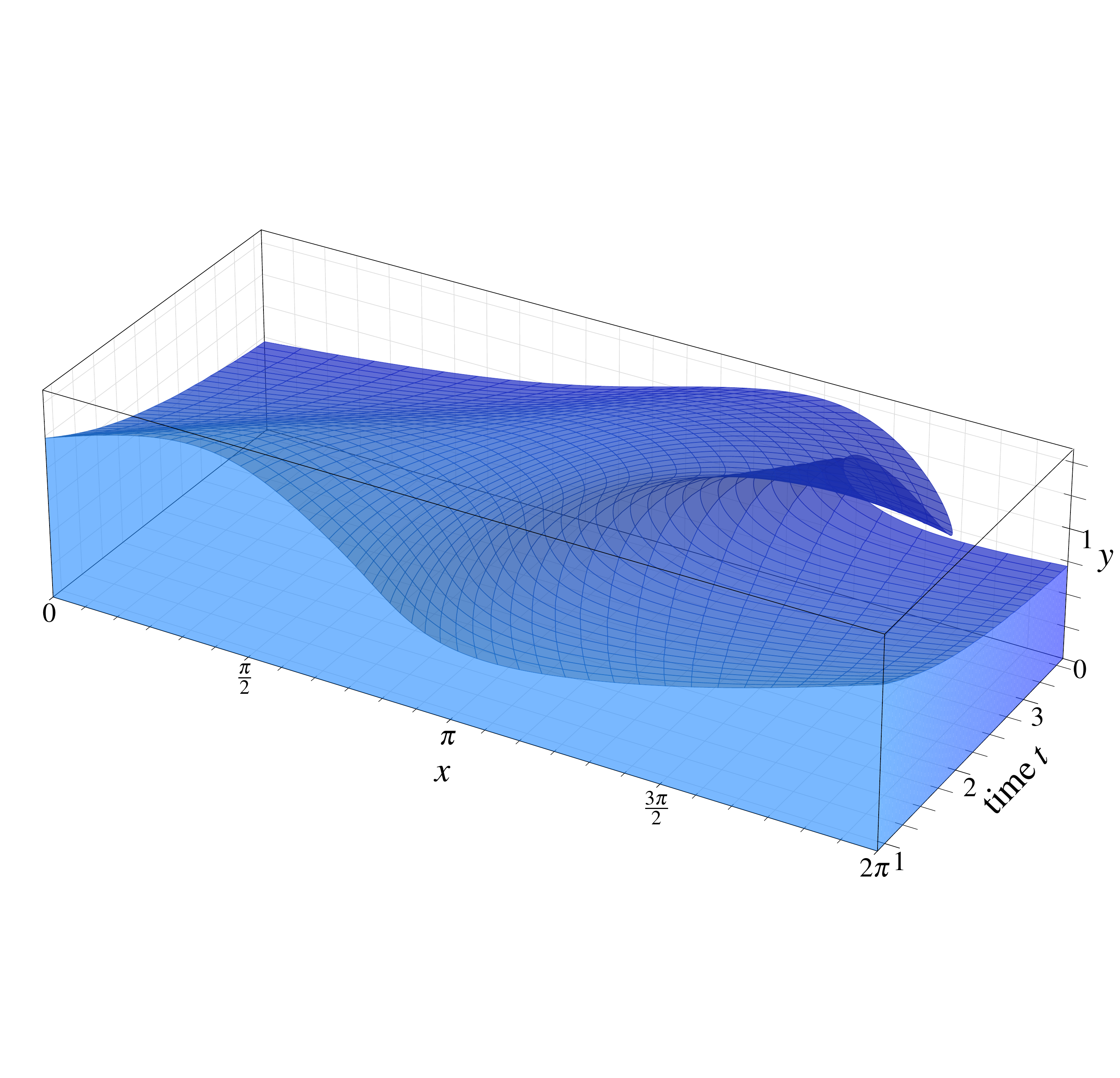}
    \caption{Evolution of the interface with time at $\mathrm{Re} = 10^6$.
     (An animation corresponding to the simulation presented in this figure is available as supplementary material.)
    }
    \label{fig:Re6Time}
\end{figure}

Even though the effects of surface tension are discarded in this study, the capillary term in \eqref{eq:NS_weak} can easily be included in the numerical scheme. Extension to two-phase flows can be considered, at the cost of meshing both domains.

Extension of this approach to the three-dimensional case, though not impossible in theory will be very challenging in practice. First because of remeshing issues in 3D, but also because the number of degrees of freedom needed to achieve convergence to such large Reynolds numbers would then become difficult to handle, even with modern supercomputers. This is due to the need of recomputing matrices at each time iteration with a moving fluid domain.
Besides, a major limitation of our approach lies in the impossibility for the interface to intersect itself. Hence the simulation has to stop as soon as a splash has occurred. Analysis of the post-breaking behavior is not permitted with our approach, but see \textit{e.g.} \citet{IAFRATI_2009, Deike2015, Deike2016, lubin_glockner_2015, Lubin2019}.

\section{The effect of viscous dissipation on the interface evolution}
\label{sec:shape}

\begin{figure}
    \centering
    \includegraphics[trim=5cm 10cm 4.5cm 11cm, clip, width=\textwidth]{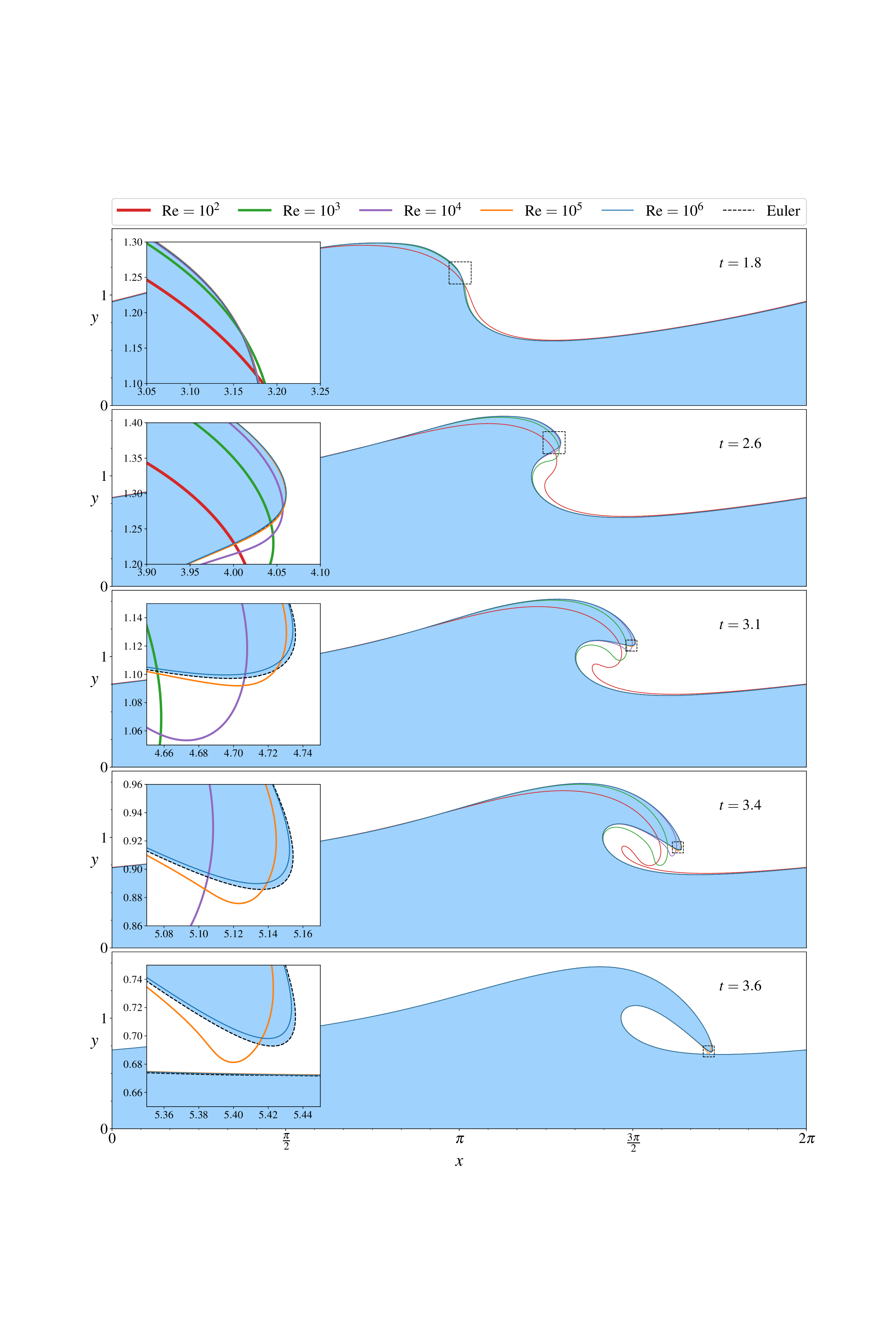}
    \caption{Interface evolution with time for different values of the Reynolds number with emphasis on the tip of the wave. The Euler solution {was obtained from} \citet{DormyLacave2023}. The shaded region corresponds to the Euler fluid domain.}
    \label{fig:ReComp}
\end{figure}

We consider numerically a domain $\Omega_{t=0}$ with $L = 2\pi$, $h_0 = 1$ and $a = 0.5$ is used. Surface tension is neglected $\sigma=0$ (thus $\text{Bo}^{-1}=0$)
in the sequel.
The domain $\Omega_t$ has been discretised with up to 4000 vertices on the water-air interface. This results in roughly $270000$ triangles and hence about $1.2$ million unknowns for the Navier-Stokes problem.

Simulations were run with Reynolds number ranging from $10^2$ to $10^6$.
Results for $\mathrm{Re} = 10^6$ are presented in figure \ref{fig:Re6Time} and a comparison of the interface for various Reynolds numbers is available in figure \ref{fig:ReComp}. The resulting wave does behave as a plunging breaker \citep[see the classification of][]{Galvin1968}. 
{We should stress that the initial condition being irrotational, the vorticity vanishes everywhere except for a thin boundary layer near the surface (see section \ref{sec:energy} below).
The fact that the vorticity is localised helps in the numerical resolution at large values of the Reynolds number.}

Our objective is now to characterize the convergence as the Reynolds number is increased. The time evolution of the interface is compared for different Reynolds number with the same initial condition in figure \ref{fig:ReComp}. The solution of the Euler equation \citep{DormyLacave2023} for the same problem is also included for comparison. The regularising effects of dissipation are clearly visible. The overhanging region takes a round shape and falls faster at larger  dissipation (\textit{i.e.} for decreasing Reynolds number).
Perhaps more surprisingly, the effects of dissipation are localised near the plunging jet. The Euler interface appears to provide a limit solution toward which the Navier-Stokes solution converges as the Reynolds number is increased.
Only a very small difference remains between the Euler solution and the Navier-Stokes solution for $\mathrm{Re} = 10^6 \, .$ This minute difference may be due to the finiteness of $\mathrm{Re}$ but also possibly to some amount of numerical diffusion as this extreme Reynolds number case is at the edge of our numerical resolution (see below).

\begin{figure}
    \begin{subfigure}{0.495\textwidth}
            \centering
    \includegraphics[width=\textwidth]{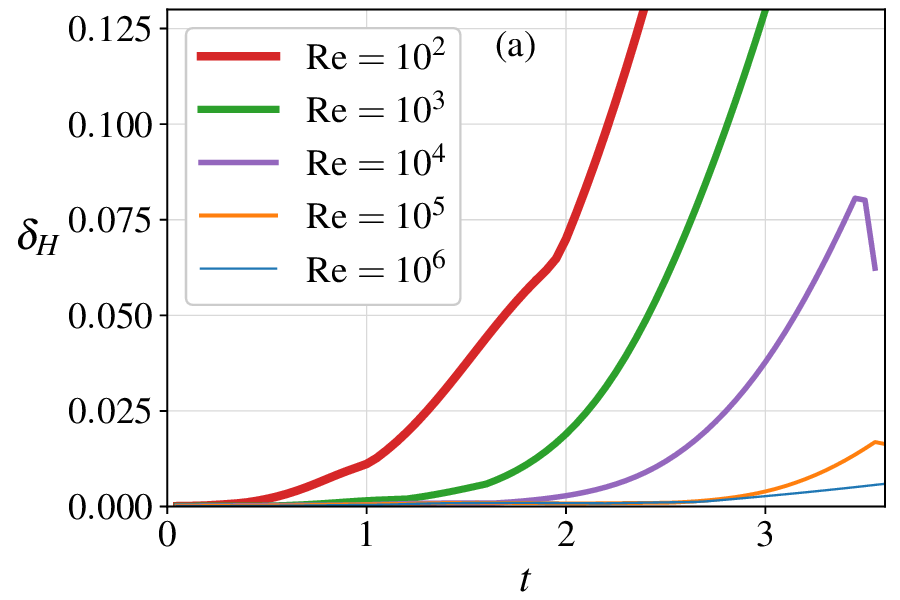}
    \end{subfigure}
    \begin{subfigure}{0.495\textwidth}
        \centering
        \includegraphics[width=\textwidth]{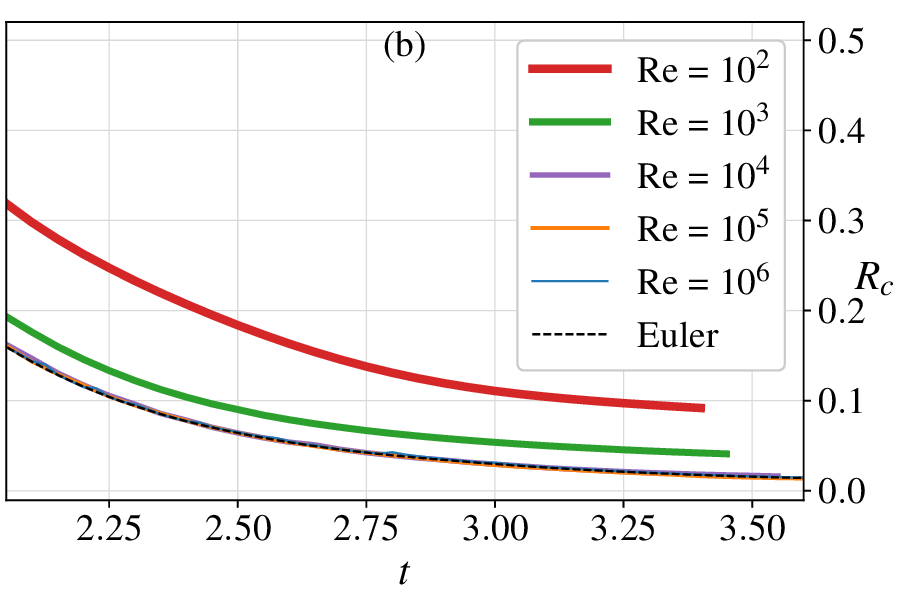}
    \end{subfigure}
    \caption{
    (a) Convergence of the Navier-Stokes solutions to the Euler 
    solution \citep{DormyLacave2023} as the Reynolds number is increased using the Hausdorff distance.
    (b)
    Time evolution of the maximum curvature radius with time for different values of the Reynolds number. The four last curves are indistinguishable at this scale.}
    \label{fig:Curv}
\end{figure}

In order to quantify the convergence of the finite Reynolds flow to the Euler solution, we must measure the differences between the various interface positions.\footnote{The numerical convergence at a given Re was assessed varying the mesh size (see Appendix A).}
We cannot use a standard norm to do that, since the interface is not a graph as soon as the wave overturns. We therefore rely \citep[as in][]{DormyLacave2023} on the bidirectional Hausdorff distance between the curves, i.e. 
\begin{equation}
\delta_H(A,B)=\max\left(\tilde{\delta}_H (A,B),\tilde{\delta}_H(B,A)\right)\, , \quad \text{with}\quad
\tilde{\delta}_H(A,B)=\max_{a\in A} \min_{b\in B} \| a-b \|  .
\end{equation}
The time evolution of the distance between each curve obtained for a given Reynolds number and the Euler solution is presented in Fig.~\ref{fig:Curv}a. The initial condition being identical, the distance is a growing function of time until the splash approaches. The time at which the effect of viscosity becomes significant increases as the Reynolds number increases.

No finite-time wedge-like singularity seems to be developing for the initial condition considered here, even in the case of the Euler solution \citep{DormyLacave2023}.
This can be assessed introducing the minimum curvature radius $R_c(t)$ (see figure~\ref{fig:Curv}b). 
The curvature of the interface can be computed numerically, with a maximum corresponding to the crest. The minimum curvature radius $R_c(t)$ is then the inverse of the maximum curvature.
Figure \ref{fig:Curv}b presents $R_c$ as a function of time. Each simulation is interrupted when the interface self-intersects. Though  $R_c$ tends to zero for large enough Reynolds number, it remains strictly positive for all time in all our simulations. No finite time singularity is obtained for this setup.
Our low Reynolds number cases $\mathrm{Re} \leq 10^3$ are characterised by a larger $R_c(t)\,.$
The fact that the curves are indistinguishable in the figure for 
$\mathrm{Re} \geq 10^4\, ,$ and coincide with the Euler simulation of \citet{DormyLacave2023}, indicate that the lack of finite time singularity for this configuration is not a consequence of viscosity
and that Bernoulli principle, accelerating the fluid near the tip of the wave \citep{pomeau2012}, 
does not cause a singularity for this initial data.
Further initial conditions and domain geometries thus need to be investigated to study the necessary conditions for the formation of such a singularity.

\section{Energy dissipation}
\label{sec:energy}

To further characterise the difference between the Euler and Navier-Stokes solutions, we now investigate the spatial distribution of viscous dissipation. A typical global energy-balance equation can be computed setting $\vect{v} = \vect{u}$ in the weak formulation \eqref{eq:NS_weak} and using the incompressibility condition,
\begin{align}
    \frac{d}{dt}\int_{\Omega_t} \frac{\vect{u}^2}{2}\,d\vect{x} &= \int_{\Omega_t} \left(\vect{g} \cdot \vect{u} - \frac{2}{\mathrm{Re}} \mathsfbi{S}(\vect{u}): \mathsfbi{S}(\vect{u}) \right)\,d\vect{x} \nonumber\\
    &{= -\frac{1}{2}\frac{d}{dt}\int_{\Gamma_{s,t}} y^2n_y\,dS - \frac{2}{\mathrm{Re}} \int_{\Omega_t} \mathsfbi{S}(\vect{u}): \mathsfbi{S}(\vect{u}) \,d\vect{x},}
\end{align}
{for a vertical gravity acceleration $\vect{g} = -\hat{\vect{y}} $ and a flat bottom. This is the usual kinetic and potential energy for water waves \citep[e.g.][]{lannes2013} with an additional viscous dissipation term. }

\begin{figure}
    \centering
    \includegraphics[width=\textwidth]{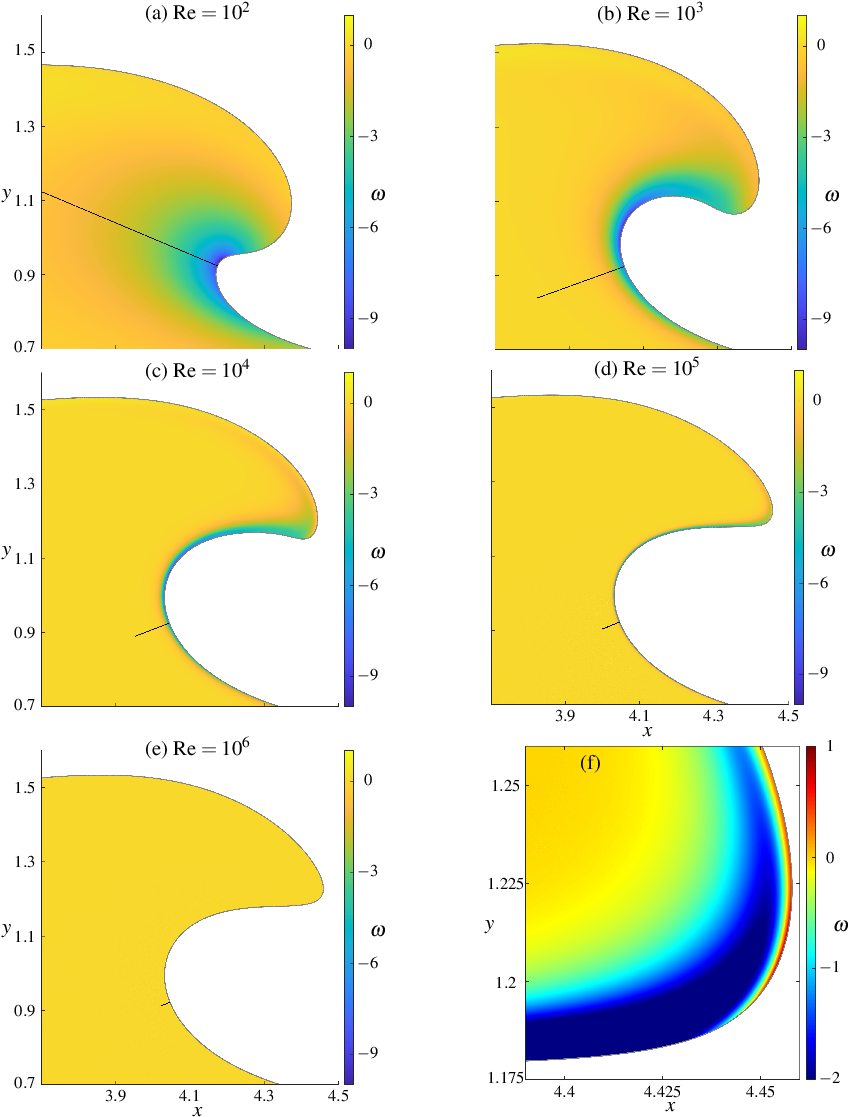}
    \caption{(a-e) Vorticity $\omega$ near the tip of the wave for different values of the Reynolds number at time $t = 2.9$. (f) A zoom on the tip of the wave for the $\mathrm{Re} = 10^5$ case (dashed rectangle in (d)). The color legend has been truncated from below to guarantee overall color coherence.}
    \label{fig:BoundaryLayer}
\end{figure}

A local equation for the kinetic energy can also be computed directly multiplying the Navier-Stokes eq. \eqref{eq:NS_normal} by $\vect{u}$ and using the typical relation $\Delta\vect{u} = -\n^\perp(\n^\perp\cdot\vect{u}) = -\n^\perp\omega$, where $\vect{u}^\perp = (-u_y,u_x)$ denotes a ${\pi}/{2}$ counter-clockwise rotation and $\omega = \n^\perp \cdot \vect{u}$ is the 2D vorticity. This leads to
\begin{equation}
    \frac{d}{dt}\frac{\vect{u}^2}{2} = \vect{g}\cdot\vect{u} - \vect{u}\cdot\n p + \frac{1}{\mathrm{Re}}\Big[\n\cdot(\vect{u}^\perp\omega)-\omega^2\Big].
\end{equation}
Hence the viscous dissipation happens in regions of the domains where the vorticity does not vanish.

\begin{figure}
    \centering
    \includegraphics[width=\textwidth]{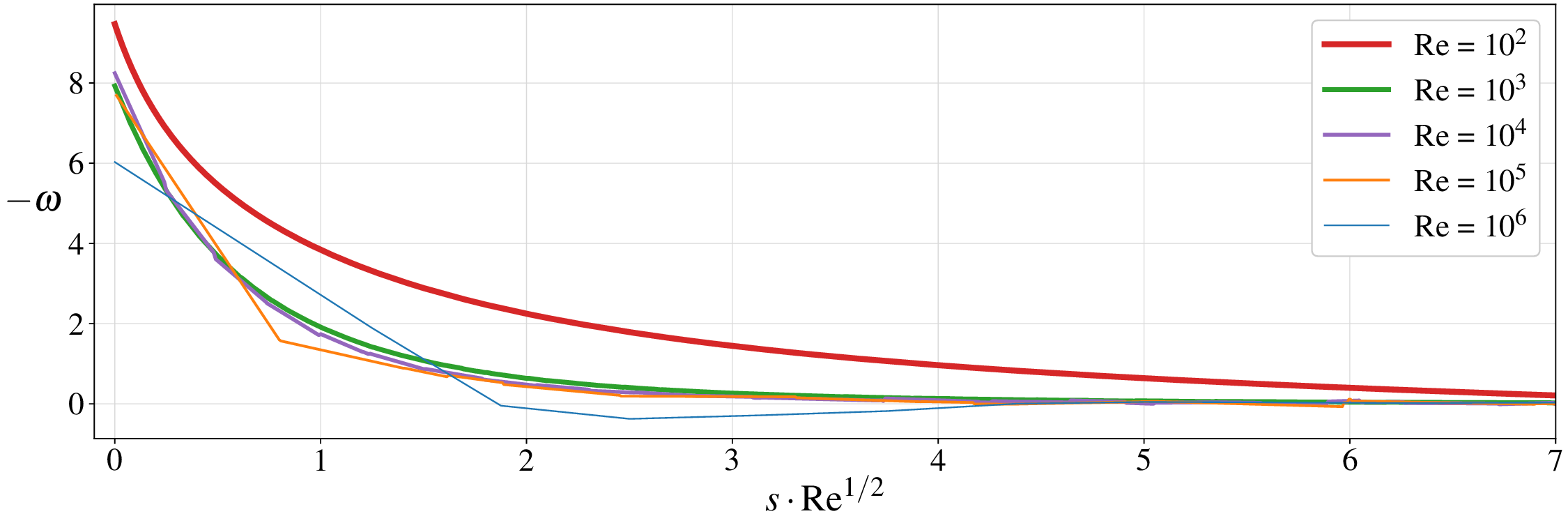}
    \caption{Vorticity cross-sections along the straight lines, normal to the boundary, shown in figure \ref{fig:BoundaryLayer} (a-e). $s$ is the arc length which parameterize the lines.}
    \label{fig:VortCuts}
\end{figure}

The vorticity is represented in figure \ref{fig:BoundaryLayer} (a-e). Note the formation of a vorticity sheet near the interface as the Reynolds number is increased. 
The \citet{LundgrenKoumoutsakos1999} theorem states that the source of mean vorticity in the boundary layer is in fact this superficial vortex sheet. It is interesting to  note that a small-magnitude positive vorticity boundary layer  appears at the tip of the wave (see Fig.~\ref{fig:BoundaryLayer}f). As time proceeds, the vorticity sheet grows where the curvature of the interface becomes important \citep[see][for the steady state case]{LonguetHiggins1992}.

The boundary layer is expected to scale as $\mathrm{Re}^{-1/2}$ (\textit{e.g.} \citet{LANDAU} for the general theory, \citet{Liu1977} for the particular case of viscous water waves {and \citet{Masmoudi2017} for a mathematical description of this limit}). We present in figure~\ref{fig:VortCuts} vorticity cross sections in the normal direction starting on the interface at $y=0.925$ (indicated on Fig.~\ref{fig:BoundaryLayer} (a-e)). 
The boundary layer of the $\mathrm{Re} = 10^6$ case is spread over 3 to 4 nodes at most, so that the exponential behavior is not clearly captured. 
Figure \ref{fig:VortCuts} nevertheless clearly illustrates the $\mathrm{Re}^{-{1}/{2}}$ scaling of the boundary layer for $\mathrm{Re} = 10^3$ to $\mathrm{Re} = 10^5$ and is compatible with the scaling for $\mathrm{Re} = 10^6$. As already mentioned, the $\mathrm{Re} = 10^2$ case is so viscous that the interface does not exhibit the same characteristics as the others.
The inward pointing normal vector at $y=0.925$ is ascending whereas the others are descending. This explains the different behavior in figure \ref{fig:VortCuts}. 
Interestingly the vorticity sheet becomes comparable in size to the minimum curvature radius $R_c$ near $\mathrm{Re} = 10^4\, ,$ i.e. when the curvature radius, as a function of the Reynolds number, reaches its minimum (figure \ref{fig:Curv}).

Another striking aspect of figure \ref{fig:VortCuts} lies in the value of the vorticity at the boundary, which appears to be fairly independent on the Reynolds number. This suggests a pointwise convergence to the interface vorticity of the Euler problem, \textit{i.e.} the vortex intensity $\gamma$ in \citet{Baker82}.

\section{The regularising effects of viscosity}

\begin{figure}
        \centering
        \begin{tikzpicture}[scale=1.5]
            \fill[color=cyan!10]
                plot [domain=-2:2,samples=200] (\x,{ - 3*sqrt(1-\x * \x/9)}) 
                -- (295:1)
                -- (295:1) arc (295:245:1)
                -- cycle;
            \draw plot [domain=-2:2,samples=200] (\x,{ - 3*sqrt(1-\x * \x/9)});
            \draw[->] (250:3) -- (250:2) node[above left]{$\hat{\vect{n}}$};
            \draw[->] (250:3) -- ++(340:1) node[below right]{$\hat{\vect{s}}$};
            \draw (280:2.15) -- (280:2.6);
            \draw (285:2.15) -- (285:2.6);
            \draw[densely dashed] (285:2.6) -- (285:3);
            \draw[densely dashed] (280:2.6) -- (280:3);
            \draw (280:3) -- (280:3.1);
            \draw (285:3) -- (285:3.1);
            \draw[<->] (280:3.1) arc (280:285:3.1) node[midway, below]{$ds$};
            \draw (277:2.25) arc (277:287:2.25);
            \draw (277:2.5) arc (277:287:2.5);
            \draw[<->] (290:3) -- (290:2.38) node[midway, right]{$n$};
            \draw[<-] (290:2.37) -- (290:1.5) node[midway, right]{$h\kappa^{-1}$} -- ++(45:0.1) -- ++(200:0.2) -- ++(45:0.125) -- (290:1.25);
            \draw[dashed] (290:1.25) -- (290:0.75) ;
            \draw[<->] (277:2.25) -- (277:2.5) node[midway, left]{$dn$};        
        \end{tikzpicture}
    \caption{Coordinate system $(s,n)$ definition and geometrical interpretation of $h$ and $\kappa$.}
    \label{fig:Reg}
\end{figure}
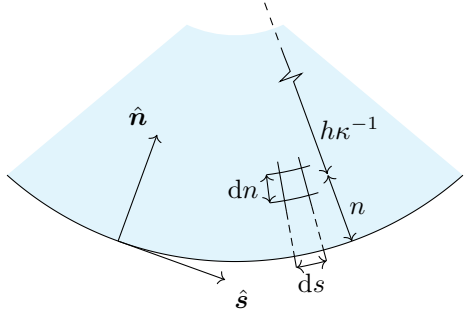

The formation of a vorticity layer near the interface is associated to the viscous regularisation of the boundary. Indeed, following \citet{LonguetHiggins1953} we can define a curvilinear coordinate system $(s,n)$ following the interface with time. This coordinate system is sometimes known as the Frenet frame. Here $s$ denotes the arc-length while $n$ is the normal coordinate, pointing inward (figure \ref{fig:Reg} (a)). We also write $\vect{u} = u_s\hat{\vect{s}} + u_n\hat{\vect{n}}\, ,$ i.e. the decomposition of the fluid velocity along the tangential and normal vectors. The time evolution of the curvature $\kappa$ (positive when convex inward) of the interface can be expressed as
\begin{equation}\label{eq:kappa}
    \frac{\partial\kappa}{\partial t} = \frac{\partial^2 u_n}{\partial s^2} - u_s\frac{\partial\kappa}{\partial s} + \kappa^2u_n,
\end{equation}
where $u_s$ and $u_n$ are evaluated at $n=0$. The full argument can be found in \citet{LonguetHiggins1953} (section 6), with a different sign convention for the curvature.

Defining the  metric factor $h = 1 - n\kappa(t,s)$ (see figure \ref{fig:Reg}), the continuity equation becomes 
\begin{equation}\label{eq:continuity_sn}
    \n\cdot\vect{u} = \partiald{u_s}{s} + \partiald{}{n}\big(hu_n\big) = 0.
\end{equation}
The full Navier-Stokes equations written in such a coordinate system can be found in \citet{LonguetHiggins1953}. We introduce the stream function $\psi$ such that 
\begin{equation}\label{eq:stream}
    u_s = \partiald{\psi}{n}\, ,\qquad\text{and}\qquad u_n = - \frac{1}{h}\partiald{\psi}{s}\,.
\end{equation}
The vorticity is defined as
\begin{equation}\label{eq:vorticity_sn}
    \omega = h^{-1}\left( \partiald{u_n}{s} - \partiald{}{n}\big(hu_s\big)\right).
\end{equation}
Where the vorticity vanishes, we can define a velocity potential $\phi$ as,
\begin{equation}
    u_s = \frac{1}{h}\partiald{\phi}{s}\qquad\text{and}\qquad u_n = \partiald{\phi}{n}.
\end{equation}
Inserting $\phi$ and $\psi$ in the continuity \eqref{eq:continuity_sn} and  vorticity \eqref{eq:vorticity_sn} equation
yields
\begin{equation}
    \Delta\phi = 0 \qquad\text{and}\qquad \Delta\psi = -\omega. 
\end{equation}
We now decompose the flow as a global potential plus a local viscous component \citep[e.g.][]{LundgrenKoumoutsakos1999},
$
\vect{u} = \n\phi + \n^\perp\psi_{\mathrm{Re}},
$
where $\n^\perp$ is defined by \eqref{eq:stream}. The  $\psi_\mathrm{Re}$ component, i.e. viscous effects, localized in a boundary layer of size $\delta = \mathrm{Re}^{-\frac{1}{2}}$, disappear as the viscosity vanishes (see Fig.~\ref{fig:BoundaryLayer}). We introduce an expansion of $\psi_\mathrm{Re}$ in powers of $\delta$,
\begin{equation}\label{eq:Psi_Expansion}
    \psi_\mathrm{Re}(s,n,t) = \delta\psi_1(s,n,t) + \delta^2\psi_2(s,n,t) + O(\delta^3).
\end{equation}
Furthermore, because of the boundary layer structure we can introduce
$
    \Psi_\mathrm{Re}\left( s,{n}/{\delta},t \right) \equiv \psi_\mathrm{Re}(s,n,t) \, .
$
Inserting the expression \eqref{eq:Psi_Expansion} into the vorticity equation \eqref{eq:vorticity_sn} leads to terms of order $O(\delta^{-1})$. Figures \ref{fig:BoundaryLayer} and \ref{fig:VortCuts} however highlight that the vorticity remains of order $O(1)$, leading to the conclusion that $\psi = O(\delta^2)$ and hence that $\psi_1 = 0$.

Rewriting the curvature evolution equation \eqref{eq:kappa} with $\phi$ and $\Psi_\mathrm{Re}$, we find out
\begin{align}\label{eq:scaling_kappa}
    \partiald{\kappa}{t} &= \frac{\partial^3\phi}{\partial n \partial^2s} - \frac{1}{h} \partiald{\phi}{s}\partiald{\kappa}{s} + \kappa^2 \partiald{\phi}{s}
    + \underbrace{\frac{\partial^2}{\partial s^2}\left( \frac{1}{h}\partiald{\psi_\mathrm{Re}}{s} \right)}_{O(\delta^2)} - \underbrace{\partiald{\kappa}{s}\partiald{\psi_\mathrm{Re}}{n}}_{O(\delta)} + \underbrace{\frac{\kappa^2}{h}\partiald{\psi_\mathrm{Re}}{s}}_{O(\delta^2)}\, .
\end{align}
The effects of viscous dissipation on the interface thus enter the leading order balance when the surface curvature becomes of order $O(\delta^{-1})$ (i.e. a curvature radius of order $O(\delta)$).
Conversely, when the surface curvature becomes of order $O(\delta)$ or smaller, viscous effects appear in time $O(\delta^{-2})\, .$
In practice, for the large Reynolds numbers applicable to water-waves, surface tension will also become significant at similar scales.

\section{Conclusions}
\label{sec:conclusions}
The present work has demonstrated that the viscous water wave problem converges toward the inviscid solution, even in the case of wave breaking. We highlighted the regularising effect of finite viscosity and quantified the curvature at which viscous effects become significant. Further work, involving different initial conditions, is needed regarding the possible formation of a finite time singularity at the tip of a breaking wave.\\[3mm]
{\bf Declaration of interest.} The authors report no conflict of interest.\\[3mm]
{\bf Acknowledgement.} The authors wish to acknowledge discussions with Christophe Lacave on water waves and with Bertrand Maury and Pierre Jolivet on the use of FreeFem.
\appendix%
\begin{figure}
      \centering
      \includegraphics[trim=0cm 0.5cm 0cm 0.5cm,clip,width=\textwidth]{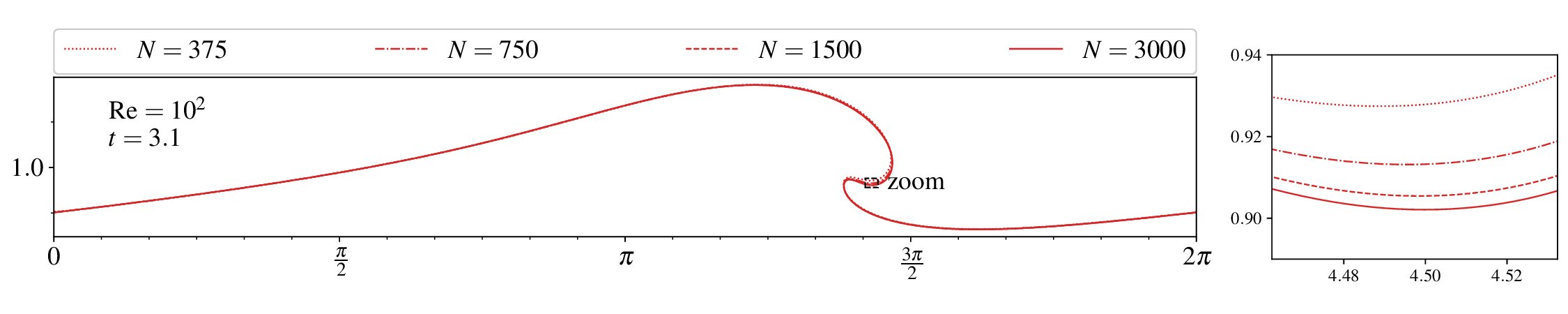}
      \includegraphics[trim=0cm 0.5cm 0cm 0.5cm,clip,width=\textwidth]{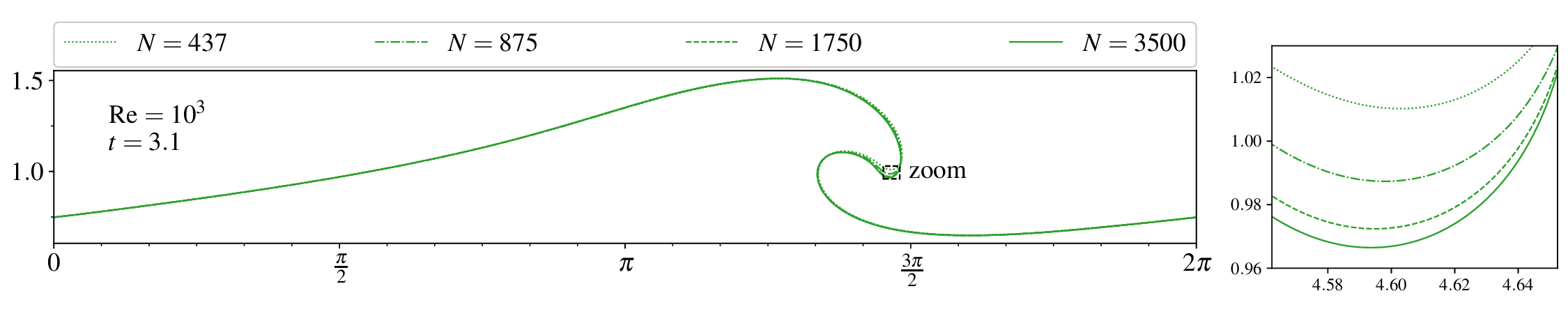}
      \includegraphics[trim=0cm 0.5cm 0cm 0.5cm,clip,width=\textwidth]{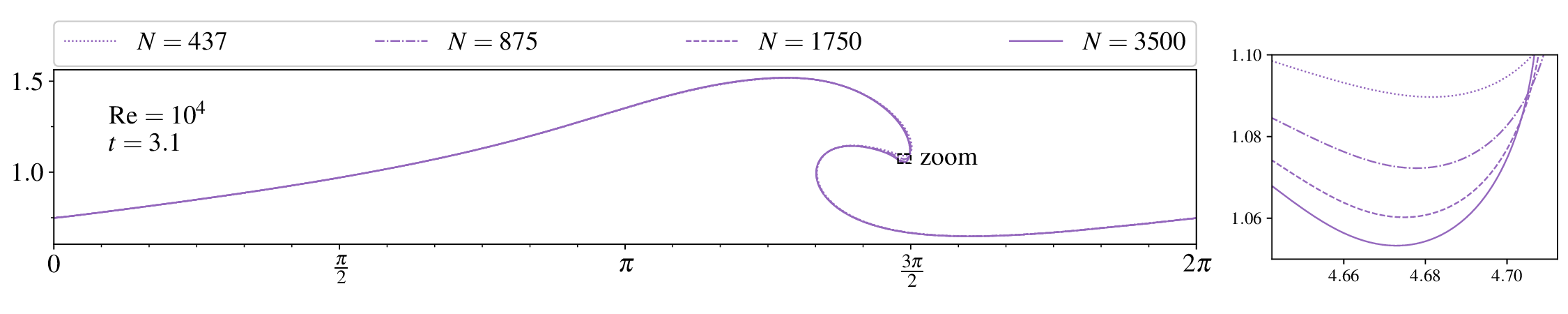}
      \includegraphics[trim=0cm 0.5cm 0cm 0.5cm,clip,width=\textwidth]{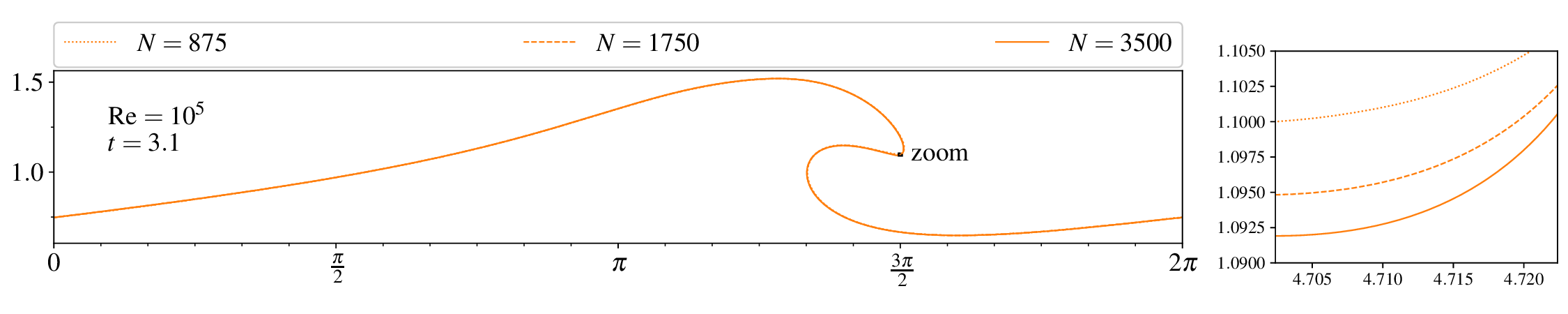}
      \includegraphics[trim=0cm 0.5cm 0cm 0.45cm,clip,width=\textwidth]{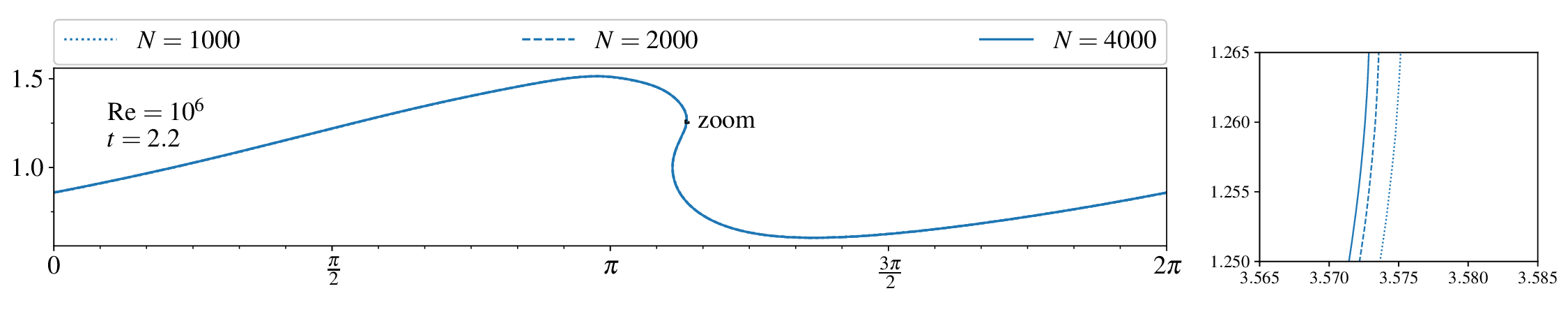}
      \includegraphics[trim=0cm 0.5cm 0cm 0.45cm,clip,width=\textwidth]{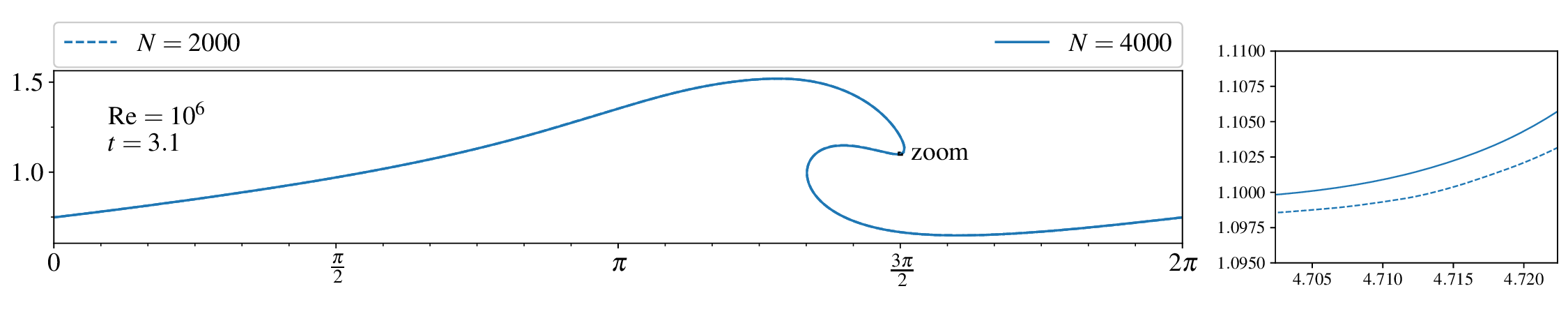}
      \caption{Numerical convergence for each Reynolds number, the rectangular region indicated on the left is blown up on the right. Note that $N=1000$ for $\mathrm{Re} = 10^6$ is unstable for times larger than $2.2$, thus only two meshes are compared in the last graph.}
      \label{fig:num_conv}
\end{figure}
{\section{Numerical convergence}
All simulations have been carried out with meshes of different sizes, measured by $N$ the number of points at the free-surface\footnote{{The total number of degrees of freedom of the finite element mesh is much larger than $N$. For example for $\mathrm{Re} = 10^6$
with $N=4000$ the final number of degrees of freedom is $\simeq 2.2 \times 10^6$.}}.
Different values of $N$ have been used depending on the Reynolds number. 
Numerical convergence is highlighted in Fig.~\ref{fig:num_conv} by using in each case grids with $N/2$, $N/4$ and $N/8$ points at the free-surface.
}

\bibliography{jfm.bib}

\end{document}